\documentclass[aps,prb,twocolumn,showpacs]{revtex4}
\usepackage{bm}
\usepackage{amsmath}
\usepackage{graphicx}

\begin{document}


\title{Pancake vortices}

\author{John R.\ Clem}
\affiliation{%
  Ames Laboratory and Department of Physics and Astronomy,\\
  Iowa State University, Ames, Iowa, 50011--3160 }

\date{\today}

\begin{abstract}
I describe the magnetic-field and current-density distributions generated
by two-dimensional (2D) pancake vortices in infinite, semi-infinite,
and finite-thickness stacks of Josephson-decoupled superconducting layers. 
Arrays of such vortices have been used to model the magnetic structure in highly
anisotropic layered cuprate high-temperature superconductors. 
I show how the electromagnetic forces between pancake vortices can be
calculatated, and I briefly discuss the effects of interlayer Josephson coupling. 
\end{abstract}

\pacs{74.25.Qt,74.72.-h,74.78.Bz,74.72.Hs}

\maketitle

\section{Introduction} 
Since this paper is intended for publication in a special Festschrift issue
honoring Mike Tinkham, I have been invited to include some personal
reflections in the introduction. I believe I first heard his name
when I was a graduate student in the early 1960s at the University of
Illinois-Urbana, working on extensions of the BCS theory\cite{Bardeen57} to
include anisotropy of the superconducting energy gap.\cite{Clem66a,Clem66b}   A
paper by Ginsberg, Richards, and Tinkham\cite{Ginsberg59} had reported results on
the far-infrared absorption in superconducting lead, which showed a precursor
hump in the real part of the complex conductivity,
$\sigma_1(\omega)/\sigma_N$.  I tried to explain this feature in
terms of gap anisotropy but was unsuccessful.

Throughout  subsequent years, I have followed Mike Tinkham's career with
considerable interest.
I have admired his research style, which consistently has resulted in new and
interesting experimental results and theoretical interpretations that advance
the theory.
I also admire anyone who can write carefully prepared books, and I
have found his books on superconductivity (in both
editions\cite{Tinkham75,Tinkham96}) to be particularly useful.
I have asked students
beginning research with me to work diligently through these books to learn the
fundamentals of superconductivity.

One of the topics that Mike Tinkham finds interesting is vortex physics, and
since this has been one of my main research interests, I
would like to focus here on one aspect: two-dimensional (2D) pancake vortices. 
This is a favorite subtopic of mine, partly because I  coined the name and partly
because my 1991 paper on this subject\cite{Clem91} has been so well received by
the superconductivity community (over 600 citations to date).  Incidentally,
although I wanted to put ``2D pancake vortex" in the title of this paper, the
editors of Physical Review B forbid this but did allow me to use these words in
the abstract and the rest of the paper.  I first reported on  my work on 2D
pancake vortices at a Gordon Research Conference chaired by Mike Tinkham in
June 1989, but (as has too often been the case with me) I was slow
to publish, and some of the key results were published in 1990 by Artemenko and
Kruglov\cite{Artemenko90} and by Buzdin and Feinberg.\cite{Buzdin90} I later
discovered that the basic solution had even been published in 1979 by
Efetov,\cite{Efetov79} but his work unfortunately had gone largely unnoticed.

This paper is organized as follows.  In Sec.\ II, I calculate the properties of 
2D pancake vortices in an infinite stack of Josephson-decoupled superconducting
layers, first by considering all the layers as being very thin and then by
considering the layers above and below the pancake layer as a
continuum.\cite{Mints00}   In Sec.\
III, I use the continuum approach to calculate the properties of  2D pancake
vortices in a semi-infinite stack of Josephson-decoupled superconducting
layers.  In Sec.\ IV, I again use the continuum approach to calculate the
properties of 2D pancake vortices in a finite stack of Josephson-decoupled 
superconducting
layers,\cite{Clem94} first for arbitrary thickness and then for a thickness much
less than the in-plane penetration depth, where the results bear some
similarities to those of Pearl\cite{Pearl64,Pearl65,deGennes66} for vortices in
thin films.  In Sec.\ V, I show how to calculate the electromagnetic forces
between pancake vortices, and in Sec.\ VI, I discuss some consequences of
Josephson coupling.  I conclude with a brief summary in Sec.\ VII.

\section{Pancake vortex in an infinite stack of superconducting layers} 

The chief motivation for my work that led to the idea of the 2D pancake vortex
was the question of how to describe the vortex structure of highly anisotropic
layered cuprate high-temperature superconductors, with Bi-2212
(Bi$_2$Sr$_2$CaCu$_2$O$_{8-\delta}$) being the best-known example.  Applying the
anisotropic Ginzburg-Landau
equations\cite{vonLaue52,Caroli63,Gorkov64,Tilley64,Tilley65,Katz69,Katz70,
Takanaka75,Klemm80,Kogan81a,Kogan81b,Kogan87,Balatskii86,Bulaevskii88} to this
material, it could easily be seen that the calculated value of the coherence
length
$\xi_c$ (the length scale describing spatial variation of the order
parameter in the
$c$ direction perpendicular to the layers) was less than the center-to-center
distance $s$ between the CuO$_2$ bilayers.  Since the Ginzburg-Landau theory
assumes that all the characteristic lengths of superconductivity are large by
comparison with atomic length scales, this fact indicated that some other theory
was needed to describe details of the vortex structure in the most anisotropic
high-T$_c$ superconductors.

The natural way to incorporate the existence of discrete layers was to make use
of the Lawrence-Doniach theory,\cite{Lawrence71} which treats the intralayer
behavior using Ginzburg-Landau theory but interlayer coupling via the Josephson
effect.\cite{Josephson62}  In this theory the coherence length $\xi_c$ plays no
role when its value is less than $s$, and the penetration depth
$\lambda_c$ describing the length scale of the spatial variation of supercurrents
parallel to the $c$ direction can be related to the maximum Josephson
supercurrent $J_0$ via\cite{Clem89}  $\lambda_c =
(c\phi_0/8\pi^2sJ_0)^{1/2}$ in Gaussian units.  The parameter
usually used to characterize the degree of anisotropy is $\gamma =
\lambda_c/\lambda_{ab}$, where $\lambda_{ab}$ is the  penetration depth
describing the length scale of the spatial variation of supercurrents
parallel to the layers (neglecting the anisotropy between the $a$ and $b$
directions, i.e., assuming for simplicity that $\lambda_a \approx \lambda_b
\approx
\lambda_{ab}$).  For Bi-2212, the value of $\gamma$ is so large that it is
difficult to measure;\cite{Kes90} $\gamma$ was found in Ref.\
\onlinecite{Martinez92} to be larger than 150, but a more recent quantitative
determination\cite{Grigorenko02} has yielded $\gamma = 640 \pm 25$.  

For such highly anisotropic materials, it seemed sensible to
me to take the limit $\gamma \rightarrow \infty$ ($\lambda_c = \infty$ or $J_0 =
0$) as the starting point to describe vortex structure.  The essential idea  was
that in a model of identical superconducting layers separated by insulating
layers, one could solve for the magnetic field and current density generated by a
2D pancake vortex in one of the superconducting layers when the 
other layers contained no vortices but served only to screen the magnetic field
generated by the pancake vortex.  With this solution as a building block, one
could then find the magnetic field produced by a stack of such pancake vortices,
even if misaligned, by the process of linear superposition.  This was basically
the approach I had used in developing the theory that quantitatively explains the
coupling forces between misaligned vortices in just two
layers,\cite{Clem74,Clem75} the primary and secondary superconducting layers of
the dc transformer  studied experimentally first by
Giaever,\cite{Giaever65,Giaever66} and later by Solomon,\cite{Solomon66}
Sherrill,\cite{Sherrill67} Deltour and Tinkham,\cite{Deltour68} and Cladis et
al.,\cite{Cladis68a,Cladis68b} but in greatest detail by Ekin et
al.\cite{Ekin74,Ekin75}

\subsection{Model of very thin discrete superconducting layers}

To calculate the magnetic-field and current-density distributions generated by a
pancake vortex in an infinite stack of Josephson-decoupled superconducting
layers, in Ref.\ \onlinecite{Clem91} I used the model in which the superconducting
layers, all of thickness
$d$, are centered on the planes $z = z_n = ns$ ($n = 0, \pm 1, \pm 2,
...$), as sketched in Fig.\ 1.
\begin{figure}
\includegraphics[width=8cm]{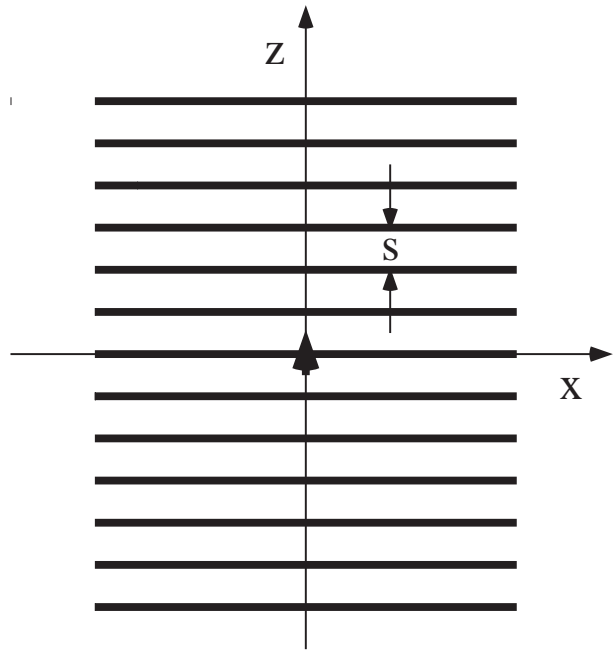}
\caption{Infinite stack of thin superconducting layers with a pancake vortex at
the origin (bold arrow).}
\label{Fig1}
\end{figure}
  The London penetration depth within each layer is
$\lambda_s$, such that the average penetration depth for currents parallel to the
layers is\cite{Clem89}
$\lambda_{||} = \lambda_s (s/d)^{1/2}$, which corresponds to the penetration
depth $\lambda_{ab}$ in the high-temperature superconductors. When the central
layer ($z = 0$) contains a vortex at the origin but all other layers are
vortex-free, the London fluxoid quantization condition\cite{London61} in layer
$n$  can be expressed as 
\begin{equation}
2 \pi \rho[a_\phi(\rho,z_n) +(2\pi \Lambda_s/c)  K_\phi(\rho,
z_n)] = \phi_0 \delta_{n0}, 
\label{London}
\end{equation} 
where in cylindrical coordinates
${\bm a}(\rho,z) = \hat \phi a_\phi(\rho,z)$ is the vector potential, ${\bm
K}(\rho,z_n) = \hat \phi K_\phi(\rho,
z_n) =  \hat \phi {\bar j_\phi}(\rho,
z_n)s  $ is the sheet-current density in layer $n$ averaged over the periodicity
length $s$, 
$\Lambda_s = 2 \lambda_{||}^2/s = 2 \lambda_s^2/d$ is the 2D screening length,
and $\phi_0 = hc/2e$ is the superconducting flux quantum.
Equation (\ref{London}) inevitably leads to a description of vortices in the
London model,\cite{Tinkham96a} which is characterized by unphysical
current-density and magnetic-field singularities on the vortex axis.  The
pioneering work on vortices by Abrikosov\cite{Abrikosov57}  showed that such
singularities are cut  off at a distance of the order of the in-plane coherence
length
$\xi_{ab}$. A simple model for the vortex core, employing a variational
core-radius parameter $\xi_v \sim \xi_{ab}$, has been used to describe straight
vortices in isotropic\cite{Clem75vm} and anisotropic\cite{Clem92}
superconductors, as well as in films of arbitrary thickness,
whether isolated\cite{Clem80} or in superconducting dc
transformers.\cite{Clem75}  This model also could be used to cure the vortex-core
singularities that are present in all the following results of this paper.

If one takes the thickness $d$ of each layer to be very small, as in Ref.\
\onlinecite{Clem91}, the vector potential  can be expressed in
the form 
\begin{equation}
a_\phi(\rho,z)=\int_0^\infty dq A(q)J_1(q\rho)Z(q,z),
\label{aexact}
\end{equation}
where $J_1(q\rho)$ is a Bessel function and $Z(q,z)$ has scallops as a function of
$z$ that are necessary to describe the discontinuities of $b_\rho(\rho,z)$ arising
from the induced sheet currents $K_\phi(\rho,z_n)$ for $n \ne 0$.  Note
that ${\bm b}(\rho,z) =
\nabla
\times {\bm a}(\rho,z)$, such that 
\begin{equation}
b_\rho(\rho,z) = -\frac{\partial a_\phi(\rho,z)}{\partial z}
\label{brho}
\end{equation}
 and
\begin{equation}
b_z(\rho,z) = \frac{1}{\rho}\frac{\partial [\rho a_\phi(\rho,z)]}{\partial \rho}.
\label{bz}
\end{equation}
  Inserting
the exact expression for
$A(q)$ into Eq.\ (\ref{aexact}) yields a complicated integral that cannot be
integrated analytically.  However, a close approximation to the exact result can
be obtained by writing $Z(q,z) =
\exp(-Q|z|)$, where $Q = (q^2+\lambda_{||}^{-2})^{1/2}$ and $A(q) =
\phi_0/ 2 \pi  \Lambda_s Q$;  this  approximation, which is valid for $s \ll
\lambda_{||}$, corresponds to retaining information on the scale of
$\lambda_{||}$ but giving up detailed
information on the finer scale of
$s$.
The resulting vector potential and magnetic field components are
\begin{eqnarray}
a_\phi(\rho,z)&=& \frac{\phi_0 \lambda_{||}}{2 \pi \Lambda_s \rho}
(e^{-|z|/\lambda_{||}}-e^{-r/\lambda_{||}}),
\label{a} \\
b_z(\rho,z)&=&\frac{\phi_0}{2 \pi \Lambda_s r}e^{-r/\lambda_{||}},
\label{bzinfty} \\
b_\rho(\rho,z)&=&\frac{\phi_0}{2 \pi \Lambda_s \rho}
\big[\frac{z}{|z|}e^{-|z|/\lambda_{||}}-\frac{z}{r}e^{-r/\lambda_{||}}\big],
\label{brhoinfty}
\end{eqnarray}
where $r = (\rho^2+z^2)^{1/2}$.  Since in the high-temperature superconductors
$s/ 2 \lambda_{||} = \lambda_{||}/\Lambda_s \approx 10^{-2}$, the vector potential
term in Eq.\ (\ref{London}) of order $\lambda_{||}/\Lambda_s$ can be neglected in
the central layer
$(n = 0)$, and we obtain to good approximation
\begin{equation}
K_\phi(\rho,z_0) = \frac{c \phi_0}{ 4 \pi^2 \Lambda_s \rho}.
\label{K0}
\end{equation}
However, for all the other layers $(n
\ne 0)$ we obtain
\begin{equation}
K_\phi(\rho,z_n) = -\frac{c \phi_0 \lambda_{||}}{ 4 \pi^2 \Lambda_s^2 \rho}
(e^{-|z_n|/\lambda_{||}}-e^{-r_n/\lambda_{||}}),
\label{Kn}
\end{equation}
where $z_n = ns$ and $r_n =  (\rho^2+z_n^2)^{1/2}$.  
Note that the magnitude
of the sheet-current density in
the $n = 0$ central layer is much
larger, by a factor of order 10$^{2}$, than the sheet-current density in
one of the  $n \ne 0$ layers. 
It is for this reason that I gave the name pancake vortex to this field and
current distribution.

An interesting property of the above solutions is that the
pancake-vortex-generated magnetic flux
$\Phi_z(\rho,z) = 2 \pi \rho a_\phi(\rho,z)$ up through a circle of radius
$\rho$ at height $z$ is (using $\Lambda_s = 2 \lambda_{||}^2/s$)
\begin{equation}
\Phi_z(\rho,z)= \phi_0 (s/2\lambda_{||})
(e^{-|z|/\lambda_{||}}-e^{-r/\lambda_{||}}),
\label{Phirho}
\end{equation}
such that the magnetic flux up through a layer at height $z$ is
\begin{equation}
\Phi_z(\infty,z)= \phi_0 (s/2\lambda_{||})
e^{-|z|/\lambda_{||}},
\label{Phiz}
\end{equation}
and the magnetic flux up through the central layer at $z=0$ is
\begin{equation}
\Phi_z(\infty,0)= \phi_0 (s/2\lambda_{||}).
\label{Phi0}
\end{equation}
When $s \ll \lambda_{||},$ as in the high-temperature superconductors, we see that
$\Phi_z(\infty,0) \ll \phi_0$.  This is at first surprising until one realizes
that fluxoids are quantized in superconductors but flux is
not.\cite{London61}  In the present problem the fluxoid is the quantity on the
left-hand side of Eq.\ (\ref{London}), and since $(2\pi
\Lambda_s/c)K_\phi(\rho,z_0)$ is proportional to $1/\rho$ and $a_\phi(\rho,z_0)$
is very small, the fluxoid is due almost entirely to the current term. Note also
that $\Phi_z(\infty,\infty) = 0$; this occurs because all the magnetic flux up
through the central layer $z = 0$ is directed radially outward by the screening
currents in the layers with $z > 0$.

On the other hand, an infinite stack of pancake vortices, whether straight or not,
has quite different magnetic-flux properties.  If there is one pancake vortex in
every layer at $z = z_n = ns$ ($n = 0, \pm 1, \pm 2,
...$), then the magnetic flux up through the central layer (and by symmetry any
other layer) is 
\begin{eqnarray}
\Phi_z(\infty,0) = \phi_0 (s/2\lambda_{||})\sum_{n=-\infty}^{\infty}
e^{-|z_n|/\lambda_{||}} = \phi_0,
\label{Phiinfty}
\end{eqnarray}
where the last equality is obtained by evaluating the sum and making use of the
property that
$s
\ll
\lambda_{||}$. Similarly, the radial magnetic field at
$\rho = \infty$ and $z = 0$ is now zero, since the positive
contributions from all the pancake vortices below the
central layer are canceled by the negative contributions from the pancake
vortices above this layer.

\subsection{Continuum model}

The solutions given in Eqs.\ (\ref{a})-(\ref{Kn}) can be obtained more
easily by regarding the $n \ne 0$ layers as a continuum, characterized by the
penetration depth $\lambda_{||}$ for currents parallel to the
layers.\cite{Mints00}     Moreover, for a
realistic treatment of stacks of just a few superconducting layers, a model
accounting for finite layer thickness $s$ is needed. We therefore use the model
sketched in Fig.\ 2
\begin{figure}
\includegraphics[width=8cm]{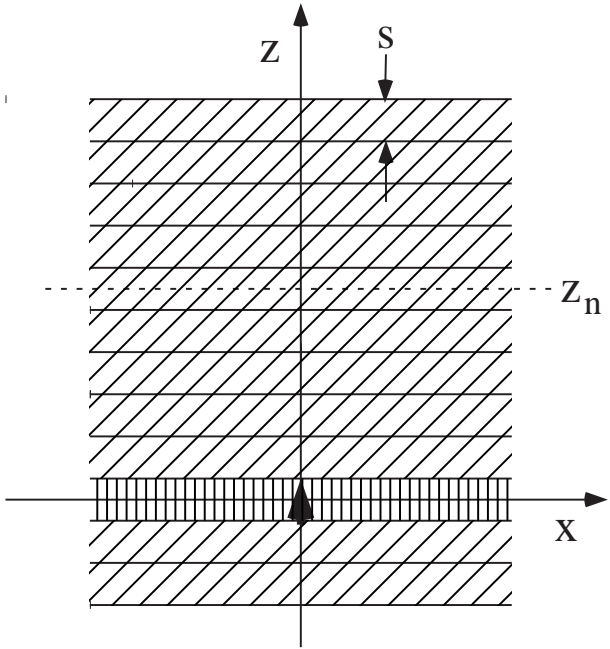}
\caption{Continuum model of an infinite stack of superconducting layers with a
pancake vortex (bold arrow at the origin) in the layer at $z = z_0 = 0$.}
\label{Fig2}
\end{figure}
and  write the London equation\cite{London61} in cylindrical
coordinates in the form
\begin{equation}
2 \pi \rho[a_\phi(\rho,z)+(4 \pi \lambda_{||}^2/c)j_\phi(\rho,z)] = \phi_0
\delta_{n0},
\label{Londonlayer}
\end{equation}
where the delta function on the right-hand side accounts for the presence of a
vortex aligned along the $z$ axis in the $n = 0$ layer ($|z| < s/2$).
Combining this equation with Ampere's law, $j_\phi = (c/4 \pi) (\partial b_\rho
/\partial z - \partial b_z / \partial \rho),$ and making use of Eqs.\
(\ref{brho}) and (\ref{bz}), we obtain the partial differential equation
\begin{equation}
\frac{\partial^2 a_\phi}{\partial z^2}+\frac{\partial^2 a_\phi}{\partial \rho^2}
+\frac{1}{\rho}\frac{\partial a_\phi}{\partial \rho}
-\big(\frac{1}{\rho^2}+\frac{1}{\lambda_{||}^2}\big)a_\phi = -\frac{\phi_0}{2 \pi
\lambda_{||}^2 \rho}\delta_{n0},
\label{aeqn}
\end{equation}
which can be solved by writing $a_\phi(\rho,z)$ in the three regions $z > s/2,
-s/2 < z <s/2,$ and $z < -s/2$ in terms of Hankel components\cite{Pearl66} as
follows:
\begin{eqnarray}
a_\phi(\rho,z) &=& \int_0^\infty dq A_a(q) J_1(q\rho)e^{-Qz}, z \ge s/2,
\label{a>cont}\\
a_\phi(\rho,z) &=&\int_0^\infty dq \big[\frac{\phi_0}{2 \pi \lambda_{||}^2
Q^2}+A_{0-}(q)e^{-Qz} \nonumber \\
&+&A_{0+}(q)e^{Qz}\big]
J_1(q\rho), -s/2 \le z \le s/2,\\
\label{apcont}
a_\phi(\rho,z) &=& \int_0^\infty dq A_b(q) J_1(q\rho)e^{-z}, z \le -s/2,
\label{a<cont}
\end{eqnarray} 
where $Q = (q^2 + 1/\lambda_{||}^2)^{1/2}.$
The four unknown functions $A_a(q), A_{0-}(q), A_{0+}(q),$ and $A_b(q)$ can be
obtained by applying the boundary conditions of continuity of $a_\phi(\rho,z)$
and $b_\rho(\rho,z)$ [Eq.\ (\ref{brho})] at the two interfaces $z = \pm s/2$,
carrying out the Hankel transforms using\cite{Jackson62} 
\begin{equation}
\int_0^\infty d\rho \rho  J_1(q\rho)J_1(q'\rho) = (1/q)\delta(q-q'),
\end{equation}
and solving the four resulting linear equations.  The results are
\begin{eqnarray}
A_{0-}(q)&=&A_{0+}(q)=-\frac{\phi_0}{4 \pi \lambda_{||}^2
Q^2} e^{-Qs/2},\\
A_a(q)&=& A_b(q) = \frac{\phi_0\sinh(Qs/2)}{2 \pi \lambda_{||}^2 Q^2}.
\end{eqnarray}
Note that $s \ll \lambda_{||}$, such that if we confine our attention to values
of $\rho \gg s$, the integrals in Eqs.\ (\ref{a>cont})-(\ref{a<cont}) are
dominated by values of $q \ll 1/s$. We  then may make the replacement
$\sinh(Qs/2)
\rightarrow Qs/2$, which makes $A_a = A_b = \phi_0/2 \pi \Lambda_s Q$, the same as
$A(q)$ in Ref.\ \onlinecite{Clem91} and Sec. II A.

The magnetic flux up through a
circle of radius $\rho$ in the plane with coordinate $z$ is $\Phi_z(\rho,z) = 2
\pi \rho a_\phi(\rho,z)$.  Evaluating the integrals for $ a_\phi(\rho,z)$ [Eqs.\
(\ref{a>cont})-(\ref{a<cont})] in the limit as $\rho
\rightarrow
\infty$,
we can show without making the approximation that $s \ll \lambda_{||}$ 
that the pancake-vortex-generated magnetic
flux through a layer at height $z$, where $|z| > s/2$,  is
\begin{equation}
\Phi_z(\infty,z)= \phi_0 \sinh(s/2\lambda_{||})
e^{-|z|/\lambda_{||}},
\label{Phizcont}
\end{equation}
and the total magnetic flux up through the central layer at $z=0$ is
\begin{eqnarray}
\Phi_z(\infty,0)= \phi_0 (1-e^{-s/2\lambda_{||}}) \approx \phi_0(s/2\lambda_{||}).
\label{Phi0cont}
\end{eqnarray}

If there is one pancake vortex in
every layer at $z = z_n = ns$ ($n = 0, \pm 1, \pm 2,
...$), even if they are misaligned, then by summing the contributions
given in Eqs.\ (\ref{Phizcont}) and (\ref{Phi0cont}) we find that the magnetic
flux up through the central layer (and by symmetry any other layer) is exactly
$\phi_0$.  If all the vortices are aligned along the $z$ axis, the magnetic-field
and current-density distributions reduce to those of the London
model,\cite{Tinkham96a} for which
\begin{eqnarray}
a_\phi(\rho) &=& \frac{\Phi_z(\rho)}{2\pi \rho} 
= \frac{\phi_0}{2\pi
\rho}\big[1-\frac{\rho}{\lambda_{||}}K_1\big(\frac{\rho}{\lambda_{||}}\big)\big],
\label{aLondon} \\
b_z(\rho)&=&\frac{\phi_0}{2\pi
\lambda_{||}^2}K_0\big(\frac{\rho}{\lambda_{||}}\big),
\label{bLondon} \\
j_\phi(\rho)&=&\frac{c \phi_0}{8 \pi
\lambda_{||}^3}K_1\big(\frac{\rho}{\lambda_{||}}\big),
\label{jLondon}
\end{eqnarray}
and $b_\rho(\rho) = 0$, where $K_n(x)$ is a modified Bessel function.  

\section{Pancake vortex in a semi-infinite stack of superconducting layers} 

In Sec. II I reviewed the results found in Ref.\ \onlinecite{Clem91} for a pancake
vortex in an infinite stack of superconducting layers, where it is seen that  the
fields and currents decay exponentially on the
scale of $\lambda_{||}$ above and below the layer containing the pancake vortex.
For a sample of thickness
$D
\gg
\lambda_{||}$ it is therefore clear that the fields and currents generated by
pancake vortices that are many $\lambda_{||}$ from either surface are essentially
the same as in Sec. II.  However, the fields and currents are
significantly altered when a pancake vortex is less than $\lambda_{||}$ 
from the surface of a sample of  thickness
$D
\gg
\lambda_{||}$ or when the sample thickness $D$ is comparable with or smaller than 
$
\lambda_{||}$.  In this section I use the continuum approximation described in
Sec.\ II B to obtain solutions describing the field and currents generated
by a vortex in an arbitrary layer of a semi-infinite stack of superconducting
layers.
  In  the next section (Sec.\ IV) I present solutions for
a pancake vortex in a finite stack of arbitrary thickness $D$.

Consider a semi-infinite stack of superconducting layers, with the top surface
on the $xy$ plane, such that all the layers are in the region $z < 0$, as
sketched in Fig\ 3.  
\begin{figure}
\includegraphics[width=8cm]{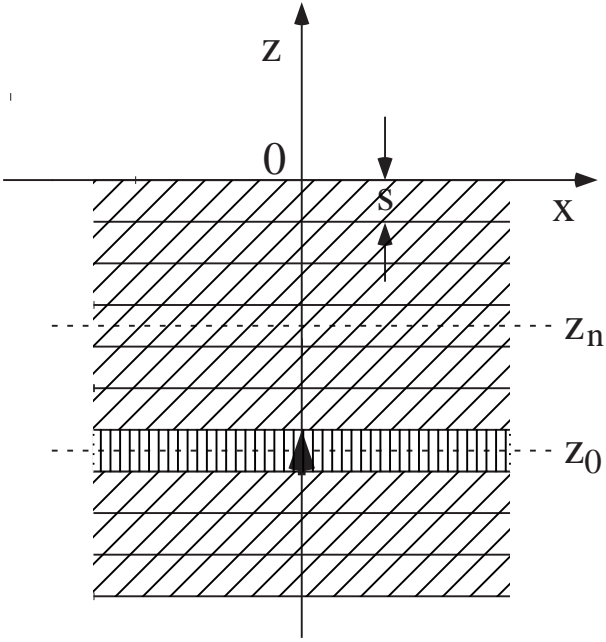}
\caption{Continuum model of a semi-infinite stack of superconducting layers
in the space $z < 0$ with a pancake vortex (bold arrow) in the layer
at
$z = z_0$.}
\label{Fig3}
\end{figure}
We  number the superconducting
layers such that the layer  $n=0$ in the region $z_{0-} < z
< z_{0+}$, centered at $z=z_0  < 0$, is the one containing the pancake
vortex.  
The other layers are centered at $z =
z_n = z_0 + ns,$  where positive (negative)
$n$ labels layers above (below) the pancake vortex.
If there are $N_+$ layers above the pancake
vortex, then the top layer is centered at $z = z_0 + N_+s = D/2-s/2.$  
By so numbering the layers, we still can
use  Eq.\ (\ref{Londonlayer}) as  the London fluxoid quantization condition. 

As in Sec.\ II, we may write the vector potential in cylindrical coordinates as
${\bm a}(\rho,z) = \hat \phi a_\phi(\rho,z)$.  However, we now have
different expressions for $a_\phi(\rho,z)$ in four regions:
\begin{eqnarray}
a_\phi(\rho,z)&=&\int_0^\infty dq A_>(q)J_1(q\rho)e^{-qz}, z  \ge 0,
\label{a>semi} \\
a_\phi(\rho,z)&=&\int_0^\infty dq
[A_{a-}(q)e^{-Q(z-z_0)} \nonumber \\
&&+A_{a+}(q)e^{Q(z-z_0)}]J_1(q\rho), \nonumber \\
&&z_{0+} \le z \le 0,
\label{aasemi} \\
a_\phi(\rho,z)&=&\int_0^\infty dq
[\frac{\phi_0}{2 \pi
\lambda_{||}^2 Q^2}
+A_{0-}(q)e^{-Q(z-z_{0})} \nonumber \\
&&+A_{0+}(q)e^{Q(z-z_{0})}]J_1(q\rho), \nonumber \\
&&
z_{0-} 
\le z
\le z_{0+},\\
\label{a0semi} 
a_\phi(\rho,z)&=&\int_0^\infty dq A_b(q) J_1(q\rho)e^{Q(z-z_0)},
z \le z_{0-},
\label{absemi}
\end{eqnarray}
where $Q = (q^2 + \lambda_{||}^{-2})^{1/2}$ and $z_{0\pm} = z_0 \pm s/2$.
The six functions $A_>(q)$, $A_{a-}(q)$, $A_{a+}(q)$, $A_{0-}(q)$, $A_{0+}(q),$
and 
$A_b(q)$, obtained by applying the six boundary conditions of continuity of
$a_\phi(\rho,z)$ and  $b_\rho(\rho,z)$ [calculated from Eq.\ (\ref{brho})] at $z =
0, z_{0+},$ and
$z_{0-}$, are
\begin{eqnarray}
A_>(q)&=&\frac{\phi_0 \sinh(Qs/2)}{\pi \lambda_{||}^2 Q^2 (1 + q/Q)}e^{Qz_0},
\label{A>semi} \\
A_{a-}(q)&=&\frac{\phi_0 \sinh(Qs/2)}{2 \pi \lambda_{||}^2 Q^2},
\label{Aa-semi} \\
A_{a+}(q)&=&\frac{\phi_0 \sinh(Qs/2)}{2 \pi \lambda_{||}^2
Q^2}\big(\frac{1-q/Q}{1+q/Q}\big)e^{2Qz_0},
\label{Aa+semi} \\
A_{0-}(q)&=&-\frac{\phi_0}{4 \pi \lambda_{||}^2 Q^2} e^{-Qs/2},
\label{A0-semi} \\
A_{0+}(q)&=&-\frac{\phi_0}{4 \pi \lambda_{||}^2
Q^2}\big[ e^{-Qs/2} \nonumber \\
&&- 2
\sinh(Qs/2)\big(\frac{1-q/Q}{1+q/Q}\big)e^{2Qz_0}\big],
\label{A0+semi} \\
A_b(q)&=&\frac{\phi_0 \sinh(Qs/2)}{2 \pi \lambda_{||}^2 Q^2}
\big[1+\big(\frac{1-q/Q}{1+q/Q}\big)e^{2Qz_0}\big].
\label{Absemi}
\end{eqnarray}

Although the resulting integrals for $a_\phi(\rho,z)$ and those [via Eqs.\
(\ref{brho}) and (\ref{bz})] for $b_\rho(\rho,z)$ and $b_z(\rho,z)$ can easily be
calculated numerically, they are too complicated to evaluate analytically for
arbitrary
$\rho$ and
$z$.  On the other hand, we can evaluate them approximately for large $\rho$. 
When
$\rho
\gg
\lambda_{||}$, the values of $q$ that dominate the integrals in Eqs.\
(\ref{a>semi})-(\ref{absemi}) via the Bessel function $J_1(q\rho)$ are those of
order
$1/\rho
\ll 1/\lambda_{||}$, such that we may replace all quantities under the integral
except $J_1(q\rho)$ by their values at
$q = 0$. Similarly, because of the factor $\exp(-qz)$  in Eq.\ (\ref{a>cont}) we
may replace
$A_>(q)$ by $A_>(0)$ to evaluate 
$a_\phi(\rho,z)$ when $\rho$ is small but
$z \gg \lambda_{||}$.

The magnetic flux up through a
circle of radius $\rho$ in the plane with coordinate $z$ is $\Phi_z(\rho,z) = 2
\pi \rho a_\phi(\rho,z)$.  Evaluating the integrals as indicated above for $
a_\phi(\rho,z)$ in the limit as $\rho
\rightarrow
\infty$, we obtain for the total magnetic flux up through the plane with
coordinate $z$:
\begin{eqnarray}
\Phi_z(\infty,z) &=& 2 \phi_0
\sinh(s/2\lambda_{||})e^{z_0/\lambda_{||}},  z \ge 0,
\label{Phiz>semi} \\
\Phi_z(\infty,z) &=&2 \phi_0
\sinh(s/2\lambda_{||})
\cosh(z/\lambda_{||})e^{z_0/\lambda_{||}},\nonumber \\
&& 
z_{0+} \le z \le 0,
\label{Phiz+0semi} \\
\Phi_z(\infty,z) &=& \phi_0
\{1-
\cosh[(z-z_0)/\lambda_{||}]e^{-s/2\lambda_{||}} \nonumber \\
&&+\sinh(s/2\lambda_{||})e^{(z+z_0)/\lambda_{||}}\}, \nonumber \\
&&z_{0-} \le z \le z_{0+},
\label{Phiz-+semi} \\
\Phi_z(\infty,z) &=&2 \phi_0
\sinh(s/2\lambda_{||})
\cosh(z_0/\lambda_{||})e^{z/\lambda_{||}}, \nonumber \\
&& z \le z_{0-}.
\label{Phizbsemi}
\end{eqnarray}

When the pancake vortex is in the top layer (i.e., when $z_0 = -s/2$), the
magnetic flux $\Phi_z(\infty,0)$ up through the top surface is approximately
$\phi_0 (s/\lambda_{||})$, since
$s/\lambda_{||} \sim 10^{-2} \ll 1$. When the pancake vortex is in a layer
much farther than $\lambda_{||}$ from the top surface, the amount of magnetic flux
up through the top surface
$\Phi_z(\infty,0)$ [Eq.\ (\ref{Phiz>semi})]
becomes exponentially small (recall that $z_0 < 0$).  The precise magnetic
field distribution generated in the space above the superconductor within
$\lambda_{||}$ of the origin can be calculated numerically for a given
pancake-vortex position
$z_0$ from Eqs.\ (\ref{brho}), (\ref{bz}), and  (\ref{a>semi}).  However, at
distances $r = \sqrt{\rho^2+z^2}$ somewhat larger than $\lambda_{||}$ from the
origin, we have to good approximation for $z \ge 0$
\begin{eqnarray}
a_\phi(\rho,z)&=&\frac{\Phi_z(\infty,0)}{2 \pi \rho} (1-\frac{z}{r}),
\\
\label{asemi}
b_\rho(\rho,z)&=& \frac{\Phi_z(\infty,0)}{2 \pi}\frac{\rho}{r^3},\\
\label{brhosemi}
b_z(\rho,z)&=& \frac{\Phi_z(\infty,0)}{2 \pi}\frac{z}{r^3}.
\label{bzsemi}
\end{eqnarray}
In other words, the magnetic field generated by the pancake vortex appears as if
generated by a magnetic monopole, with the flux $\Phi_z(\infty,0)$ [Eq.\
(\ref{Phiz>semi})] spreading out evenly into the hemisphere above the surface.
It is important to note that only pancake vortices within about $\lambda_{||}$
(or $\lambda_{ab}$ in the high-temperature superconductors) are visible using
Bitter decoration, scanning Hall-probe microscopy, scanning SQUID microscopy, or
magneto-optical techniques; pancake vortices deeper than this make an
exponentially small contribution to the magnetic field above the surface. 

From Eq.\ (\ref{Phiz-+semi}) we see that the magnetic flux
up through the plane $z = z_0$ in the layer containing the
pancake vortex is
\begin{eqnarray}
\Phi_z(\infty,z_0)&=&\phi_0[1-e^{-s/2\lambda_{||}}
+\sinh(s/2\lambda_{||})e^{2z_0/\lambda_{||}}] \nonumber \\
&\approx& \phi_0(s/2\lambda_{||})(1+e^{2z_0/\lambda_{||}}).
\label{Phisemiz0}
\end{eqnarray}
When the pancake vortex is in the top layer (i.e., if $z_0 = -s/2$), the magnetic
flux up through this layer is approximately $\phi_0 
(s/\lambda_{||})$, and when the pancake vortex is deep inside the
superconductor  (i.e., if $-z_0 \gg \lambda_{||}$), the magnetic  flux up
through the pancake layer is approximately
$\phi_0  (s/2\lambda_{||})$, as found in Sec.\ II for the infinite
superconductor [Eqs.\ (\ref{Phi0}) and (\ref{Phi0cont})]. 

If there is a pancake vortex in every layer, even if they are misaligned, the
total magnetic flux up through any plane with coordinate $z$ is exactly equal to
$\phi_0$.  This can be shown by replacing $z_0$ by $z_n = z_0 +ns$, noting that
the top layer is centered at $-s/2$,  and  summing over all $n$, using Eq.\
(\ref{Phiz>semi}) if
$z>0$.  On the other hand, if
$z<0$, one must use Eq.\
(\ref{Phizbsemi}) for the top layers for which $z_n - s/2 \ge z$,  
Eq.\ (\ref{Phiz-+semi}) for the layer containing $z$ for which $z_n - s/2 \le z
\le z_n + s/2$, and  
Eq.\ (\ref{Phiz+0semi}) for the remaining layers for which 
$z_n + s/2 \le z$.  If all the pancake vortices are aligned along the $z$ axis, 
the magnetic-field and current-density distributions 
reduce to those calculated by Pearl\cite{Mints00,Pearl66,Clem70} for a vortex in a
semi-infinite superconductor. 

Scanning Hall-probe experiments visualizing vortices in underdoped,
highly anisotropic YBa$_2$Cu$_3$O$_{6+x}$ (YBCO) single crystals, where $x =
0.35-0.375$, recently have been carried out by Guikema.\cite{Guikema04}  In the
most underdoped crystals, the observations revealed what at first appeared to be
``partial vortices" carrying magnetic flux less than $\phi_0$. Guikema 
concluded, however,  that  such images are caused by a full vortex that is
partially displaced horizontally, i.e., a ``split pancake-vortex stack." The
magnetic flux generated above the surface by the two parts of the vortex stack
can be calculated as follows.  Suppose the  bottom portion, consisting of
pancake vortices below the plane $z = -d$, is aligned along the
$z$ axis,  and  the top portion,
consisting of pancake vortices  above the plane
$z = -d$, is aligned parallel to the $z$ axis but at $(x,y) = (x_0,0)$.
Using Eq.\ (\ref{Phiz>semi}) to sum the contributions from the pancake vortices
in the two portions, one finds that the magnetic flux $\Phi_{bot} = \phi_0
\exp(-d/\lambda_{||})$ generated by the bottom portion emerges from the vicinity
of the origin
$(x,y,z) = (0,0,0)$, and  the magnetic flux $\Phi_{top} = \phi_0
[1-\exp(-d/\lambda_{||})]$ generated by the top portion emerges from the
vicinity of the point
$(x,y,z) = (x_0,0,0)$.  The two flux contributions should be resolvable when the
displacement
$x_0$ exceeds the Hall-probe size and the probe's field sensitivity allows detection of
both contributions.

\section{Pancake vortex in a finite stack of superconducting layers} 

Since all laboratory samples of the high-temperature superconductors are of
finite thickness, it is important to examine how the properties of
pancake vortices discussed in Sec. II are modified when we take the finite
thickness into account, including the possibility that the thickness $D$ may be
less than the penetration $\lambda_{||}$.  Let us begin by considering a pancake
vortex centered on the
$z$ axis in a stack of superconducting layers, each of thickness $s$, in the
region
$-D/2 < z < D/2$, as sketched in Fig.\ 4.
\begin{figure}
\includegraphics[width=8cm]{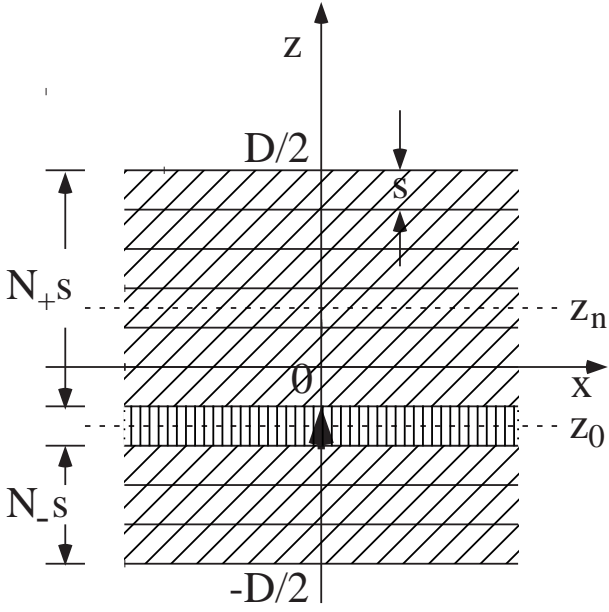}
\caption{Continuum model of a stack of superconducting layers
in the space $|z| < D/2$ with a pancake vortex (bold arrow) in the layer
at
$z = z_0$.}
\label{Fig4}
\end{figure}
As in the previous section, we number
the layers such that the layer $n = 0$ at $z = z_0$, where $|z_0| \le (D/2-s/2),$
is the one containing the pancake vortex.  The other layers are centered at $z =
z_n = z_0 + ns,$  where positive (negative)
$n$ labels layers above (below) the pancake vortex.
If there are $N_+$ layers above the pancake
vortex, then the top layer is centered at $z = z_0 + N_+s = D/2-s/2,$  and if
there are
$N_-$ layers below the pancake vortex, then the bottom layer is centered at $z  =
z_0 - N_- s= -D/2+s/2.$
As in Sec. II B, I treat all the layers
using the continuum approximation and 
use  Eq.\ (\ref{Londonlayer}) as  the London fluxoid quantization condition.
In Sec. IV A, I show how to calculate the fields generated by a
pancake vortex in a finite stack of Josephson-decoupled superconducting layers,
each of thickness $s$, with an arbitrary total stack thickness $D$ relative to
$\lambda_{||}$.  In Sec. IV B, I consider the simplifications that arise when $D
\ll  \lambda_{||}$, which corresponds to the case of high-temperature
superconducting samples consisting of roughly ten or fewer unit cells along the
$c$ direction.

\subsection {Finite stack of arbitrary thickness}

As in Secs.\ II B and III, we  write the vector potential in cylindrical
coordinates as
${\bm a}(\rho,z) = \hat \phi a_\phi(\rho,z)$.  However, we now have
different expressions for $a_\phi(\rho,z)$ in five regions:
\begin{eqnarray}
a_\phi(\rho,z)&=&\int_0^\infty dq A_>(q)J_1(q\rho)e^{-q(z-D/2)}, \nonumber \\ && z
\ge D/2,
\label{aD>} \\
a_\phi(\rho,z)&=&\int_0^\infty dq
[A_{a-}(q)e^{-Q(z-D/2)}\nonumber \\ && +A_{a+}(q)e^{Q(z-D/2)}]J_1(q\rho),
\nonumber \\ && z_{0+}
\le z \le D/2,
\label{aDa-+} \\
a_\phi(\rho,z)&=&\int_0^\infty dq
[\frac{\phi_0}{2 \pi
\lambda_{||}^2 Q^2}+A_{0-}(q)e^{-Q(z-z_{0})}\nonumber \\ &&
+A_{0+}(q)e^{Q(z-z_{0})}]J_1(q\rho),
\nonumber \\
&&z_{0-} \le z \le z_{0+},
\label{aD0-+} \\
a_\phi(\rho,z)&=&\int_0^\infty dq 
[A_{b-}(q)e^{-Q(z+D/2)}\nonumber \\ &&
+A_{b+}(q)e^{Q(z+D/2)}]J_1(q\rho),\nonumber \\ &&  -D/2 \le  z \le z_{0-},
\label{aDb-+} \\
a_\phi(\rho,z)&=&\int_0^\infty dq A_<(q) J_1(q\rho)e^{q(z+D/2)},\nonumber \\ && 
z \le -D/2,
\label{aD<}
\end{eqnarray}
where $Q = (q^2 + \lambda_{||}^{-2})^{1/2}$, the subscript $a$ ($b$) denotes
the layered region above (below)  the pancake vortex,  and $z_{0\pm} = z_0 \pm
s/2$.  The eight functions
$A_>(q)$, $A_{a-}(q)$, $A_{a+}(q)$, $A_{0-}(q)$, $A_{0+}(q)$, $A_{b-}(q)$,
$A_{b+}(q),$ and 
$A_<(q)$, obtained by applying the eight boundary conditions of continuity of
$a_\phi(\rho,z)$ and $b_\rho(\rho,z)$ [calculated from Eq.\
(\ref{brho})] at $z = D/2, z_{0-}, z_{0+}$, 
and $-D/2$, are
\begin{eqnarray}
A_>(q)&=&\frac{\phi_0}{\pi  \lambda_{||}^2 Q^2} \sinh(Qs/2) G(q,z_0),
\label{A>D} \\
A_{a-}(q)&=&\frac{\phi_0}{2 \pi  \lambda_{||}^2 Q^2} \sinh(Qs/2) (1+q/Q) G(q,z_0),
\label{Aa-D} \\
A_{a+}(q)&=&\frac{\phi_0}{2 \pi  \lambda_{||}^2 Q^2} \sinh(Qs/2) (1-q/Q)
G(q,z_0),
\label{Aa+D} \\
A_{0-}(q)&=&-\frac{\phi_0}{4 \pi  \lambda_{||}^2 Q^2}
[e^{Qs/2} \nonumber \\ &-&  2 \sinh(Qs/2)(1+q/Q)G(q,z_0)e^{QD/2}e^{-Qz_0}]
\nonumber \\& = &\frac{\phi_0}{4 \pi  \lambda_{||}^2 Q^2}
[e^{-Qs/2} \nonumber \\&-&2
\sinh(Qs/2)(1-q/Q)G(q,-z_0)e^{-QD/2}e^{-Qz_0}],\nonumber \\
\label{A0-D} \\
A_{0+}(q)&=&-\frac{\phi_0}{4 \pi  \lambda_{||}^2 Q^2}
[e^{-Qs/2} \nonumber \\ &-& 2 \sinh(Qs/2)(1-q/Q)G(q,z_0)e^{-QD/2}e^{Qz_0}]
\nonumber \\ &=&-\frac{\phi_0}{4 \pi  \lambda_{||}^2 Q^2}[e^{Qs/2} \nonumber \\
&-&2 \sinh(Qs/2)(1+q/Q)G(q,-z_0)e^{QD/2}e^{Qz_0}] \nonumber \\
\label{A0+D} \\
A_{b-}(q)&=&\frac{\phi_0}{2 \pi  \lambda_{||}^2 Q^2} \sinh(Qs/2) (1-q/Q)
G(q,-z_0),
\label{Ab-D} \\
A_{b+}(q)&=&\frac{\phi_0}{2 \pi  \lambda_{||}^2 Q^2} \sinh(Qs/2) (1+q/Q)
G(q,-z_0),
\label{Ab+D} \\
A_<(q)&=&\frac{\phi_0}{\pi  \lambda_{||}^2 Q^2} \sinh(Qs/2) G(q,-z_0),
\label{A<D}
\end{eqnarray}
where
\begin{equation}
G(q,z)=\frac{(1+q/Q)+(1-q/Q)e^{-QD}e^{-2Qz}}{(1+q/Q)^2-(1-q/Q)^2e^{-2QD}}
e^{-QD/2}e^{Qz}.
\label{G}
\end{equation}

Although the resulting integrals for $a_\phi(\rho,z)$ and those [via Eqs.\
(\ref{brho}) and (\ref{bz})] for $b_\rho(\rho,z)$ and $b_z(\rho,z)$ can easily be
calculated numerically, they are too complicated to evaluate analytically for
arbitrary
$\rho$ and
$z$.  On the other hand, we can evaluate them approximately for large $\rho$. 
When
$\rho
\gg
\lambda_{||}$, the values of $q$ that dominate the integrals in Eqs.\
(\ref{aD>})-(\ref{aD<}) via the Bessel function $J_1(q\rho)$ are those of
order
$1/\rho
\ll 1/\lambda_{||}$, such that we may replace all quantities under the integral
except $J_1(q\rho)$ by their values at
$q = 0$. Similarly, because of the factors $\exp(-qz)$ and $\exp(qz)$ in Eqs.\
(\ref{aD>}) and (\ref{aD<}) we may replace
$A_>(q)$ by $A_>(0)$ and $A_<(q)$ by $A_<(0)$ to evaluate 
$a_\phi(\rho,z)$ when $\rho$ is small but
$|z|-D/2 \gg \lambda_{||}$.

The magnetic flux up through a
circle of radius $\rho$ in the plane with coordinate $z$ is $\Phi_z(\rho,z) = 2
\pi \rho a_\phi(\rho,z)$.  Evaluating the integrals as indicated above for $
a_\phi(\rho,z)$ in the limit as $\rho
\rightarrow
\infty$, we obtain for the total magnetic flux up through the plane with
coordinate $z$:\cite{Clem94}
\begin{widetext}
\begin{eqnarray}
\Phi_z(\infty,z) &=& 2 \phi_0
\sinh\big(\frac{s}{2\lambda_{||}}\big)
\cosh\big(\frac{D/2+z_0}{\lambda_{||}}\big)
/\sinh\big(\frac{D}{\lambda_{||}}\big), 
z \ge D/2,
\label{Phiz>D} \\
\Phi_z(\infty,z) &=&2 \phi_0
\sinh\big(\frac{s}{2\lambda_{||}}\big)
\cosh\big(\frac{D/2+z_0}{\lambda_{||}}\big) 
\cosh\big(\frac{D/2-z}{\lambda_{||}}\big)
/\sinh\big(\frac{D}{\lambda_{||}}), 
z_{0+} \le z \le D/2,
\label{Phiz+D} \\
\Phi_z(\infty,z) &=& \phi_0
\{1-[e^{(D-s)/2\lambda_{||}}\cosh\big(\frac{D/2-z_0}{\lambda_{||}}\big)
-e^{-(D-s)/2\lambda_{||}}\cosh\big(\frac{D/2+z_0}{\lambda_{||}}\big)]
e^{z/\lambda_{||}}/2\sinh\big(\frac{D}{\lambda_{||}}\big)
\nonumber \\
&&-[e^{(D-s)/2\lambda_{||}}\cosh\big(\frac{D/2+z_0}{\lambda_{||}}\big)
-e^{-(D-s)/2\lambda_{||}}\cosh\big(\frac{D/2-z_0}{\lambda_{||}}\big)]
e^{-z/\lambda_{||}}/2\sinh\big(\frac{D}{\lambda_{||}}\big)\}, \nonumber \\
&&z_{0-} \le z \le z_{0+},
\label{Phiz-+D} \\
\Phi_z(\infty,z) &=&2 \phi_0
\sinh\big(\frac{s}{2\lambda_{||}}\big)
\cosh\big(\frac{D/2-z_0}{\lambda_{||}}\big)
\cosh\big(\frac{D/2+z}{\lambda_{||}}\big)
/\sinh\big(\frac{D}{\lambda_{||}}), 
-D/2 \le z \le z_{0-},
\label{Phiz-D} \\
\Phi_z(\infty,z) &=& 2 \phi_0
\sinh\big(\frac{s}{2\lambda_{||}}\big)
\cosh\big(\frac{D/2-z_0}{\lambda_{||}}\big)
/\sinh\big(\frac{D}{\lambda_{||}}\big),  
z \le -D/2.
\label{Phiz<D}
\end{eqnarray}
\end{widetext}

The magnetic flux $\Phi_z(\infty,D/2)$ up through the top surface is
given by Eq.\ (\ref{Phiz>D}).  When $D \gg \lambda_{||}$ and a pancake vortex is 
in the top layer (i.e., when $z_0 = D/2-s/2$),  we obtain $\Phi_z(\infty,D/2)
\approx \phi_0 (s/\lambda_{||}),$ which is a tiny fraction of $\phi_0$,
since
$s/\lambda_{||} \sim 10^{-2} \ll 1$. As a function of the distance $D/2-z_0$ of
the pancake vortex from the top surface, we see that $\Phi_z(\infty,D/2)
\approx \phi_0 (s/\lambda_{||})\exp{[-(D/2-z_0)/\lambda_{||}]}.$  
When $D \ll \lambda_{||},$ we find that  $\Phi_z(\infty,D/2)
\approx \phi_0 (s/D) =  \phi_0/N,$ independent of the position $z_0$ of the
pancake vortex within the stack, where $N = D/s$ is the number of layers in the
sample.  When $N = D/s = 1$, $\Phi_z(\infty,D/2) = \phi_0$, because our
results then reduce to those of Pearl,\cite{Pearl64,Pearl65,deGennes66} who
calculated the field and current distribution generated by a vortex in a film
of thickness  much less than the London penetration depth.  The precise
magnetic field distribution generated in the space above the superconductor  can
be calculated numerically for a given
$z_0$ from Eqs.\ (\ref{brho}), (\ref{bz}), and  (\ref{aD>}).  However, at
distances $r_+ = \sqrt{\rho^2+(z-D/2)^2}$ from the
point on the surface directly above the pancake vortex that are larger than
$\lambda_{||}$ when $D > 2\lambda_{||}$ or larger than the two-dimensional
screening length
$\Lambda_D = 2\lambda_{||}^2/D$ when $D < 2\lambda_{||}$, we have to good
approximation for
$z
\ge D/2$
\begin{eqnarray}
a_\phi(\rho,z)&=&\frac{\Phi_z(\infty,D/2)}{2 \pi \rho}
\big[1-\frac{(z-D/2)}{r_+}\big],
\\
\label{aD}
b_\rho(\rho,z)&=& \frac{\Phi_z(\infty,D/2)}{2 \pi}\frac{\rho}{r_+^3},\\
\label{brhoD}
b_z(\rho,z)&=& \frac{\Phi_z(\infty,D/2)}{2 \pi}\frac{(z-D/2)}{r_+^3}.
\label{bzD}
\end{eqnarray}
In other words, the magnetic field generated by the pancake vortex appears as if
generated by a positive magnetic monopole, with the flux $\Phi_z(\infty,D/2)$
[Eq.\ (\ref{Phiz>D})] spreading out into the hemisphere above the surface.

Similar statements can be made about the magnetic flux $\Phi_z(\infty,-D/2)$ up
through the bottom surface [Eq.\ (\ref{Phiz<D})].  At large distances  $r_- =
\sqrt{\rho^2+(z+D/2)^2}$ from the point on the surface directly below the pancake
vortex, the magnetic field appears as if generated by a negative magnetic
monopole.

From Eq.\ (\ref{Phiz-+D}) we see that the magnetic flux
up through the plane $z = z_0$ in the layer containing the
pancake vortex is
\begin{eqnarray}
\Phi_z(\infty,z_0)&=&\phi_0\{1-[\sinh\big(\frac{D-s/2}{\lambda_{||}}\big)
\nonumber
\\&-&\sinh\big(\frac{s}{2\lambda_{||}}\big)\cosh\big(\frac{2z_0}{\lambda_{||}}\big)]
/\sinh\big(\frac{D}{\lambda_{||}}\big)\}.\nonumber \\
\label{PhiDz0}
\end{eqnarray}
When $D \gg \lambda_{||}$, the dependence of this magnetic flux upon the
distance $(D/2 - |z_0|)$ from the top or bottom surface is given by 
$\Phi_z(\infty,z_0) \approx \phi_0 (s/2\lambda_{||})
\{1+\exp{[-2(D/2-|z_0|)/\lambda_{||}}]\}.$  When 
the pancake vortex is in
the top or bottom layer (i.e., if
$|z_0| = D/2-s/2$), the magnetic flux up through this layer is approximately
$\phi_0  (s/\lambda_{||})$, and when the pancake vortex is deep inside the
superconductor  (i.e., if $D/2-|z_0| \gg \lambda_{||}$), the magnetic  flux up
through the pancake layer is approximately
$\phi_0  (s/2\lambda_{||})$, as found in Sec.\ II for the infinite
superconductor [Eqs.\ (\ref{Phi0}) and (\ref{Phi0cont})]. 
When $D \ll \lambda_{||},$ we see that $\Phi_z(\infty,z_0) \approx  \phi_0 (s/D)
=  \phi_0/N,$ independent of the position $z_0$ of the pancake vortex within the
stack, where $N = D/s$ is the number of layers in the sample. 

If there is a pancake vortex in every layer, even if they are misaligned, the
total magnetic flux up through any plane with coordinate $z$ is exactly equal to
$\phi_0$.  This can be shown by replacing $z_0$ by $z_n = z_0 +ns$  and  summing
over all $n$, using Eq.\ (\ref{Phiz>D}) if
$z>D/2$ or Eq.\ (\ref{Phiz<D}) if
$z<-D/2$.  On the other hand, if
$|z|<D/2$, one must use Eq.\
(\ref{Phiz-D}) for the top layers for which $z_n - s/2 \ge z$,  
Eq.\ (\ref{Phiz-+D}) for the layer containing $z$ for which $z_n - s/2 \le z
\le z_n + s/2$, and  
Eq.\ (\ref{Phiz+D}) for the remaining layers for which 
$z_n + s/2 \le z$.
If all the vortices are aligned along the $z$ axis, the magnetic-field and
current-density distributions reduce to those given in Ref.\ \onlinecite{Clem80}
when $\xi_v = 0$.

It is possible that 
scanning Hall-probe or magneto-optical experiments may be able to
detect  partial
vortices or split pancake-vortex stacks\cite{Guikema04} carrying magnetic flux
less than $\phi_0$ in samples of highly anisotropic layered
superconductors of thickness
$D <
\lambda_{||}$.   The magnetic flux
generated above the surface $z = D/2$ by the two parts of the vortex stack can be
calculated as follows.  Suppose the  bottom portion, consisting of pancake
vortices below the plane $z = D/2-d$, is aligned along the
$z$ axis,  and  the top portion,
consisting of pancake vortices  above the plane
$z = D/2-d$, is aligned parallel to the $z$ axis but at $(x,y) = (x_0,0)$.
Using Eq.\ (\ref{Phiz>D}) to sum the contributions from the pancake vortices
in the two portions, one finds that the magnetic flux $\Phi_{bot} = \phi_0
\sinh[(D-d)/\lambda_{||}]/\sinh(D/\lambda_{||})$ generated by the bottom
portion emerges from the vicinity of the point
$(x,y,z) = (0,0,D/2)$, and  the magnetic flux $\Phi_{top} = \phi_0
\{1-\sinh[(D-d)/\lambda_{||}]/\sinh(D/\lambda_{||})\}$ generated by the top
portion emerges from the vicinity of the point
$(x,y,z) = (x_0,0,D/2)$.  The two flux contributions should be resolvable when the
displacement
$x_0$ exceeds the Hall-probe size and the probe's field sensitivity allows detection of
both contributions. Note that $\Phi_{bot} = \Phi_{top} = \phi_0/2$ when $d = D/2
\ll \lambda_{||}$.

\subsection {Finite stack of thickness $D \ll \lambda_{||}$}

Considerable simplifications occur when the thickness
$D = Ns$ of the stack is much less than the in-plane penetration depth
$\lambda_{||}$.\cite{Mints00}  It is well known from the work of Refs.\
\onlinecite{Pearl64} and \onlinecite{Pearl65}  that when $D \ll \lambda_{||}$ the
characteristic screening length  in isolated films is not $\lambda_{||}$ but
rather the 2D screening length
$\Lambda_D = 2\lambda_{||}^2/D$.  This is also true for the case of
Josephson-decoupled stacks of total thickness $D$ considered here.  We may derive
equations for
$a_\phi(\rho,z),$ $b_\rho(\rho,z),$ and $b_z(\rho,z)$ valid for $D \ll
\lambda_{||}$ and  $\rho \gg \lambda_{||}$ by starting with Eqs.\
(\ref{aD>})-(\ref{aD<}), applying Eqs.\ (\ref{brho}) and (\ref{bz}), and making
the replacement
$e^{\pm Qz} = \cosh(Qz) \pm
\sinh(Qz)$.  Since we are most interested in values of $\rho$ of the order of
$\Lambda_D$ or larger, because of the presence of $J_1(q\rho)$ the dominant values
of $q$ in the resulting integrals are of the order of $q \sim 1/\Lambda_D \ll
1/\lambda_{||}$, such that
$Q$ can be replaced by $1/\Lambda_{||}$, and small quantities of the order of
$D/\lambda_{||}$ and $q\lambda_{||}$ are of the same order of magnitude. 
Expanding in powers of the small quantities ($D/\lambda_{||}$ and
$q\lambda_{||}$), we find that both $a_\phi(\rho,z)$ and $b_z(\rho,z)$ are
to lowest order independent of $z$, with small correction terms of the order of
$D/\lambda_{||}$, such that to good approximation  we may write these quantities
as
\begin{eqnarray}
a_\phi(\rho,z)&=&\frac{\phi_0}{2 \pi N}\int_0^\infty dq
\frac{J_1(q\rho)}{1+q\Lambda_D}e^{-q(z-D/2)}, z
\ge D/2,\nonumber \\
\label{aDthin>} \\
a_\phi(\rho,z)&=&\frac{\phi_0}{2 \pi N}\int_0^\infty dq
\frac{J_1(q\rho)}{1+q\Lambda_D}, -D/2 \le z \le D/2,\nonumber \\
\label{aDthin0} \\
a_\phi(\rho,z)&=&\frac{\phi_0}{2 \pi N}\int_0^\infty dq
\frac{J_1(q\rho)}{1+q\Lambda_D}e^{q(z+D/2)},
z \le -D/2,\nonumber \\
\label{aDthin<} \\
b_z(\rho,z)&=&\frac{\phi_0}{2 \pi N}\int_0^\infty dq
\frac{qJ_0(q\rho)}{1+q\Lambda_D}e^{-q(z-D/2)}, z
\ge D/2,\nonumber \\
\label{bzDthin>} \\
b_z(\rho,z)&=&\frac{\phi_0}{2 \pi N}\int_0^\infty dq
\frac{qJ_0(q\rho)}{1+q\Lambda_D}, -D/2 \le z \le D/2,\nonumber \\
\label{bzDthin0} \\
b_z(\rho,z)&=&\frac{\phi_0}{2 \pi N}\int_0^\infty dq
\frac{qJ_0(q\rho)}{1+q\Lambda_D}e^{q(z+D/2)}.
z \le -D/2,\nonumber \\
\label{bzDthin<}
\end{eqnarray}
On the other hand, the radial component of the magnetic field varies strongly
with $z$:
\begin{widetext}
\begin{eqnarray}
b_\rho(\rho,z)&=&\frac{\phi_0}{2 \pi N}\int_0^\infty dq
\frac{qJ_1(q\rho)}{1+q\Lambda_D}e^{-q(z-D/2)}, z
\ge D/2,
\label{brhoDthin>}\\
b_\rho(\rho,z)&=&\frac{\phi_0}{2 \pi N}\int_0^\infty dq
\frac{qJ_1(q\rho)}{1+q\Lambda_D}, z = D/2,
\label{brhoDthinD/2} \\
b_\rho(\rho,z)&=&b_\rho(\rho,D/2)
+\frac{(D/2-z)}{(D/2)}\frac{a_\phi(\rho)}{\Lambda_D}, z_{0+} \le z \le D/2,
\label{brhoDthin+}\\
b_\rho(\rho,z)&=&b_\rho(\rho,D/2)
+\frac{(D/2-z)}{(D/2)}\frac{a_\phi(\rho)}{\Lambda_D}
-\frac{(z_{0+}-z)}{(D/2)}\frac{\phi_0}{2 \pi \Lambda_D\rho},
z_{0-} \le z \le z_{0+},
\label{brhoDthin0} 
\end{eqnarray}
\begin{eqnarray}
b_\rho(\rho,z)&=&b_\rho(\rho,D/2)
+\frac{(D/2-z)}{(D/2)}\frac{a_\phi(\rho)}{\Lambda_D}
-\frac{s}{(D/2)}\frac{\phi_0}{2 \pi \Lambda_D\rho}, 
-D/2 \le z \le
z_{0-},
\label{brhoDthin-}\\
b_\rho(\rho,z)&=&b_\rho(\rho,D/2)
-\frac{2}{\Lambda_D}\big[\frac{\phi_0}{2 \pi N \rho}-a_\phi(\rho)\big]
\nonumber \\
&=&b_\rho(\rho,-D/2),  z = -D/2,
\label{brhothin-D/2}\\
b_\rho(\rho,z)&=&-\frac{\phi_0}{2 \pi N}\int_0^\infty dq
\frac{qJ_1(q\rho)}{1+q\Lambda_D}e^{q(z+D/2)},
z \le -D/2,
\label{brhothin<}
\end{eqnarray}
\end{widetext}
where we use  $a_\phi(\rho)$ to denote the vector potential in the region $|z| \le
D/2$, since
$a_\phi(\rho,z)$ is very nearly independent of $z$.
The sheet current $K_n(\rho) = K_\phi(\rho,z_n) = sj_\phi(\rho,z_n)$ in layer
$n$ can be obtained  from either $j_\phi(\rho,z)=(c/4\pi)\partial
b_\rho(\rho,z)/\partial z$ or  the fluxoid quantization condition [Eq.\
(\ref{Londonlayer})]:
\begin{equation}
K_n(\rho) = \frac{c}{2\pi \Lambda_s} \big[\frac{\phi_0}{2 \pi \rho} \delta_{n0}
-a_\phi(\rho)\big],
\label{Knthin}
\end{equation}
where $\Lambda_s = 2 \lambda_{||}^2/s = N \Lambda_D$.
The net sheet current through the thickness $D$ is the sum of the $K_n$:
\begin{eqnarray}
K_D(\rho) = \sum_{n=-N_-}^{N_+} K_n(\rho) = 
\frac{c}{2\pi \Lambda_D} \big[\frac{\phi_0}{2 \pi N \rho} 
-a_\phi(\rho)\big].
\label{KDthin}
\end{eqnarray}

The integrals appearing in Eqs.\ (\ref{aDthin>})-(\ref{KDthin}), which are
evaluated in Appendix A, have simple forms in the limits $D \ll \rho \ll
\Lambda_D$ and $\rho \gg \Lambda_D$. The corresponding expressions for
the physical quantities we have calculated in this section are given in Table
\ref{1}. 
The magnetic-field and current-density distributions reduce to the thin-film
results of Pearl\cite{Pearl64,Pearl65} when $N=1$ and $D=s$ or when each of the
$N$ layers contains a pancake vortex on the $z$ axis.

\begin{table}
\centering
\caption{\label{1}Results for one pancake vortex in a stack of $N$ superconducting
layers of total thickness $D = Ns \ll \lambda_{||}$ in the limits $D \ll \rho \ll
\Lambda_D = 2\lambda_{||}^2/D$ and $\rho
\gg
\Lambda_D$.  Since $D$ is very small, $r=(\rho^2+z^2)^{1/2}$ may be regarded as
the distance from the pancake vortex, and $|z|$ may be regarded as the distance
from the top or bottom surface.}
\begin{ruledtabular}
\begin{tabular}{ccc}
{\bf Physical quantity} & $\bm {\rho \ll \Lambda_D}$ & $\bm
{\rho
\gg \Lambda_D}$\\
$a_\phi(\rho,z)$ & $\frac{\phi_0(r-|z|)}{2\pi N \Lambda_D \rho}$ & 
$\frac{\phi_0(r-|z|)}{2\pi N \rho r}$\\
$a_\phi(\rho,0)$ & $\frac{\phi_0}{2\pi N \Lambda_D}$ & $\frac{\phi_0}{2\pi N
\rho}$\\
$\Phi_z(\rho,z) = 2 \pi \rho a_\phi(\rho,z)$ & $\frac{\phi_0 (r-|z|)}{N\Lambda_D}$
&
$\frac{\phi_0(r-|z|)}{Nr}$\\
$\Phi_z(\rho,0) = 2 \pi \rho a_\phi(\rho,0)$ & $\frac{\phi_0 \rho}{N\Lambda_D}$ &
$\frac{\phi_0}{N}$\\
$b_\rho(\rho,z)$, $z=\pm |z|$ & $\pm \frac{\phi_0(r-|z|)}{2\pi N \Lambda_D \rho
r}$ & 
$\pm
\frac{\phi_0 \rho}{2\pi N r^3}$\\
$b_\rho(\rho,\pm D/2)$ & $\pm \frac{\phi_0}{2\pi N \Lambda_D \rho}$ & $\pm
\frac{\phi_0}{2\pi N
\rho^2}$\\
$b_z(\rho,z)$ & $\frac{\phi_0}{2\pi N \Lambda_D r}$ & $\frac{\phi_0 z}{2\pi N
r^3}$\\
$b_z(\rho,0)$ & $\frac{\phi_0}{2\pi N \Lambda_D \rho}$ & $\frac{\phi_0
\lambda_D}{2\pi N
\rho^3}$\\
$K_0(\rho)= K_\phi(\rho,z_0)$ & $\frac{c \phi_0}{4\pi^2 N \Lambda_D \rho}$ &
$\frac{c
\phi_0 (N-1)}{4\pi^2 N^2 \Lambda_D \rho}$\\
$K_n(\rho)= K_\phi(\rho,z_n)$, $n \ne 0$ & $-\frac{c \phi_0}{4\pi^2 N^2
\Lambda_D^2}$ &
$-\frac{c
\phi_0 }{4\pi^2 N^2 \Lambda_D \rho}$\\
$K_D(\rho)$ & $\frac{c \phi_0}{4\pi^2 N \Lambda_D \rho}$ & $\frac{c \phi_0
}{4\pi^2 N \rho^2}$
\end{tabular}
\end{ruledtabular}
\end{table}

\section{Forces}

The force on a second pancake vortex at the position $(\rho,z_n)$ due to a
pancake vortex centered on the $z$ axis at $(0,z_0)$ can be calculated from the
Lorentz force.\cite{Tinkham96b}  Since  pancake vortices cannot move out of their
planes, the force is directed parallel to the planes in the radial $(\hat \rho)$
direction:
\begin{equation}
F_\rho(\rho) = K_\phi(\rho,z_n) \phi_0 /c,
\end{equation}
where 
\begin{equation}
K_\phi(\rho,z_n) = \frac{c}{2\pi \Lambda_s} \big[\frac{\phi_0}{2 \pi \rho}
\delta_{n0} -a_\phi(\rho,z_n)\big]
\label{Kphi}
\end{equation}
is the sheet-current density and $a_\phi(\rho,z_n)$ is the vector
potential at $(\rho,z_n)$ generated by the pancake vortex at $(0,z_0)$, and
$\Lambda_s = 2
\lambda_{||}^2/s = N \Lambda_D$.

If both pancake vortices are in the same plane, the interaction force is
always repulsive and in an infinite or semi-infinite stack of
superconducting layers is given to excellent approximation by 
\begin{equation}
F_\rho(\rho) = \frac{\phi_0^2}{4\pi^2\Lambda_s \rho}
\label{F0}
\end{equation}
for all $\rho$. The reason for this is that the vector potential in Eq.\
(\ref{Kphi}) obeys
$a_\phi(\rho,z_0) \le (s/\lambda_{||})(\phi_0/2\pi\rho)
\ll
\phi_0/2\pi\rho$, as shown in Secs. II
and III.  However, for a finite stack of thickness $D \ll \lambda_{||}$ consisting
of
$N$ layers, Eq.\ (\ref{F0}) holds only for small
$\rho$ ($\rho \ll \Lambda_D$), where the vector potential in Eq.\
(\ref{Kphi}) is much smaller than $\phi_0/2\pi\rho$.  As
discussed in Sec.\ IV A, the magnetic flux at infinite radius $\Phi_z(\infty,z_0)$
up through the pancake-vortex layer is approximately
$\phi_0/N$, which means that $a_\phi(\rho,z_0) \approx \phi_0/2 \pi N \rho$ for
large
$\rho$, and   
\begin{equation}
F_\rho(\rho) = \frac{(N-1)}{N} \frac{\phi_0^2}{4\pi^2 \Lambda_s \rho},
\rho \gg \Lambda_D.
\label{FN}
\end{equation}
In the special case when $N = 2$, the repulsive force given in Eq.\ (\ref{FN})
is half that in Eq.\ (\ref{F0}). 

If the two pancake vortices are in different planes, the  $\phi_0/2\pi\rho$ term
in Eq.\ (\ref{Kphi}) is absent, and the interaction force is given by 
\begin{equation}
F_\rho(\rho) = -\frac{\phi_0 a_\phi(\rho,z_n)}{2\pi\Lambda_s}.
\label{Fn}
\end{equation}
Because $a_\phi(\rho,z_n)$ is always positive, the interaction force is always
negative, i.e., in a direction so as to cause the two pancake vortices to become
aligned along the same vertical axis.  
For the general case, it is not a simple matter to calculate the
spatial dependence of the attractive force between pancake vortices in different
layers, as can be seen from the expressions for
$a_\phi(\rho,z)$ given in previous sections.  However, we can say that for an
infinite or semi-infinite stack of superconducting layers, the magnitude of this
attractive force is orders of magnitude smaller than the repulsive force between
pancake vortices in the same layer.  The attractive force between vortices in
different layers in an infinite stack (or deep
inside a semi-infinite stack) has a range $\lambda_{||}$ in the $z$ direction.
Equation (\ref{a}) shows that the attractive force in the infinite stack vanishes
exponentially when the interplanar separation of the pancakes along the $z$
direction exceeds
$\lambda_{||}$.  For a finite
stack of thickness
$D = Ns \ll \lambda_{||}$, we find that the attractive force between pancake
vortices in different layers is
\begin{equation}
F_\rho(\rho) = -\frac{\phi_0^2}{4\pi^2 \Lambda_s^2}
=-\frac{\phi_0^2}{4\pi^2 N^2 \Lambda_D^2}, D \ll \rho \ll
\Lambda_D,
\label{FnD<}
\end{equation}
which agrees with the force in the infinite stack calculated from Eq.\ (\ref{Kn})
when
$|z_n|
\ll \rho \ll
\lambda_{||}$, and
\begin{equation}
F_\rho(\rho) 
=-\frac{\phi_0^2}{4\pi^2 N^2 \Lambda_D \rho},  \rho \gg
\Lambda_D.
\label{FnD>} 
\end{equation}
For the special case of two layers $(N= 2)$ and a separation $\rho \gg \Lambda_D$,
the magnitude of the attractive force exerted by a pancake vortex in one layer
upon a pancake vortex in the other layer [Eq.\ (\ref{FnD>})] is equal to the
magnitude of the repulsive force between two pancake vortices in the same layer
[Eq.\ (\ref{FN}) with
$\Lambda_s = N \Lambda_D$]. 

The energy per unit length of a uniformly tilted infinite stack of pancake
vortices in an infinite stack of superconducting layers was calculated in Ref.\
\onlinecite{Clem91}.  The corresponding line tension $T(\theta)$ was calculated 
in Ref.\
\onlinecite{Benkraouda96} as a function of the angle $\theta$
relative to the $z$ axis and found to be positive only for $\theta < 51.8^\circ
$, indicating an instability beginning at $51.8^\circ$.  Further
calculations\cite{Benkraouda96} showed that, because pancake vortices
energetically prefer to line up parallel to the $z$ axis, the energy for an
infinite stack of pancake vortices with a large average tilt angle is
reduced when the stack breaks up into shorter stacks parallel to the $z$ axis with
kinks between them.  Pe et al.\cite{Pe97a} calculated the equilibrium positions
of a stack of pancake vortices in a finite stack of Josephson-decoupled layers
when equal and opposite transport currents are applied to the top and bottom
layers.  They found that the pancake vortices in the top and bottom layers have
large displacements to the left and right, while the other vortices all remain
close to the $z$ axis.  Related model calculations were carried out in Ref.\
\onlinecite{Pe97b} for moving two-dimensional
pancake vortex lattices in a finite stack of magnetically coupled superconducting
thin films with transport current only in the top layer.  For small currents, the
entire electromagnetically coupled vortex array moves uniformly in the
direction of the Lorentz force but with a large displacement of the pancake
vortices in the top layer relative to the others, which remain in nearly
straight lines perpendicular to the layers.  Above a critical decoupling current,
the 2D vortex array in the top layer periodically slips relative to the arrays in
the other layers, and the dc current-voltage characteristics for the top and
bottom layers resemble those reported  in Refs.\
\onlinecite{Ekin74} and \onlinecite{Ekin75} for the dc transformer.

\section{Josephson coupling}

The equations underlying the solutions presented in Secs.\ II, III, and IV assume
no interlayer Josephson coupling.  Implicit in these solutions is the assumption
that the component of the magnetic field parallel to the
layers spreads out uniformly in the radial direction.  This is consistent with
the idea that if a magnetic field is applied parallel to a stack of
Josephson-decoupled layers, the field will penetrate uniformly between the
layers.  

When the layers are Josephson-coupled, however, parallel magnetic
fields penetrate the structure in the form of quantized Josephson
vortices.\cite{Bulaevskii73,Clem90}  As discussed in Ref.\
\onlinecite{Clem90}, Josephson vortices in the high-temperature superconductors
have highly elliptical current and field patterns.  Since the decay length for
currents perpendicular to the layers is
$\lambda_c$ and that for currents parallel to the layers is $\lambda_{ab}$, 
the ratio of the width of the
pattern parallel to the layers to the height perpendicular to the layers is
$\gamma =
\lambda_c/\lambda_{ab}$ at
large distances from the non-linear Josephson core.  For  a high-$\kappa$
Abrikosov vortex\cite{Abrikosov57} in an isotropic superconductor, the decay
length at large distances is the penetration depth $\lambda$, and the currents in
the nonlinear core vary on the much smaller length scale of the coherence length
$\xi$.  The behavior in a Josephson vortex is
analogous.  The small length scale for spatial variation of the Josephson currents
in the vortex core (whose axis is centered in the insulating layer between two
adjacent superconducting layers) is the Josephson
length\cite{Clem90,Bulaevskii92a}
$\lambda_J =
\gamma s$, while the corresponding length scale for the return of these currents
parallel to the layers is $s$, such that the ratio of the width to the height
of the Josephson core is 
$\gamma =
\lambda_J/s = \lambda_c/\lambda_{ab}$.

In the presence of interlayer Josephson coupling,  the
magnetic-field and current-density distributions generated by a pancake vortex
are unaltered at short distances but are strongly affected at distances of the
order of $\lambda_J$ and $\lambda_c$.  To give an example, imagine an infinite
stack of semi-infinite Josephson-coupled superconducting layers, all parallel to
the
$xy$ plane, filling the half-space $x>0$, such that the surface coincides with
the plane $x = 0$.  Imagine creating a
pancake vortex at the origin in the superconducting layer $n = 0$ and moving it
in to a distance $x_0$.  The magnetic-field and current-density distributions,
including the effects of a dipole-like stray field that leaks out into the space
$x<0$, have been calculated as a function of $x_0$  in Ref.\ \onlinecite{Mints96}
under the assumption of very weak Josephson coupling.   In the presence of
Josephson coupling, however, the component of the magnetic field parallel to the
layers cannot penetrate with a power-law dependence to large distances but rather
must decay exponentially with the decay length
$\lambda_c$, because this component of the field induces Josephson currents to
flow perpendicular to the layers. As the pancake vortex moves deeper into the
stack, the Josephson coupling  begins to play a greater role. When the pancake
vortex is a distance 
$\lambda_J$ or greater from the surface, a Josephson core region of width
$2 \lambda_J$ appears in the region between the vortex axis and the surface.
Finally, at distances  such that 
$x_0 \gg \lambda_c$, the current and field distribution can be characterized as a
pancake vortex in which the fields at distances less than $\lambda_J$ from the
axis are nearly the same as in the Josephson-decoupled case, and the magnetic
flux carried up through the pancake layer $z = 0$ is $\phi_0(s/2\lambda_{||})$. 
However, this magnetic flux does not flow radially outward to infinity as in the
Josephson-decoupled case but instead is confined within a highly
elliptical field distribution consisting of an overlapping Josephson
vortex-antivortex pair, which links the pancake vortex to the surface. 
Recall that when a straight vortex is at a distance $x$ from the surface of an
isotropic superconductor of penetration depth $\lambda$, the magnetic flux inside
the superconductor, calculated accounting for the image vortex at $-x$, is
$\phi_0 [1-\exp(-x/\lambda)]$.  As a pancake vortex moves from the surface
to a position $x_0$ deep within the superconductor, it drags along a Josephson
vortex (carrying magnetic flux in the $+x$ direction) whose axis is in the
insulating layer at $z = -s/2$, and it also drags along a Josephson antivortex
(carrying magnetic flux in the $-x$ direction) whose axis is in the insulating
layer at $z = +s/2$.  Accounting for the overlapping field distributions, which
nearly cancel each other, we find that the magnetic flux carried in the $+x$
direction through the space $z<0$ is $\phi_0 [1-\exp(-s/2\lambda_{||})] \approx
\phi_0(s/2\lambda_{||})$; the same amount of magnetic flux is carried back in the
$-x$ direction through the space $z > 0$.

To give another example, consider an infinite stack of pancake vortices initially
aligned along the $z$ axis in an infinite stack of
Josephson-decoupled superconducting layers.  As discussed at the end of Sec.\ II,
the field and current distributions reduce to those of a line vortex in an
isotropic superconductor of penetration depth
$\lambda_{||}$.\cite{Tinkham96a}  The magnetic field is everywhere perpendicular
to the layers. Now imagine displacing all of the  pancake vortices  in the space
$z > s/2$ by a distance $x_0$ in the $x$ direction, such that the pancake vortex
stack now has a kink at $z = s/2$.  In the absence of Josephson coupling, the
resulting field and current distributions can be obtained by superposing those
given in Sec.\ II.  A component of the field parallel to the layers must arise in
order to displace the magnetic flux $\phi_0$ whose distribution is centered on
the $z$ axis for $z \ll -\lambda_{||}$ to a distribution centered on the line
$(x,y) = (x_0,0)$ for $z \gg \lambda_{||}$.   The component of the field parallel
to the layers  has a dipole-like distribution in any plane $z
=$ const, with a power-law dependence at large distances, but it decreases
exponentially for $|z| >
\lambda_{||}$ because of the screening currents that flow parallel to the layers.
In the presence of interlayer Josephson coupling, the above picture is altered,
and it is  now useful to think of kinked vortices as  stacks of pancake
vortices connected by Josephson strings (short pieces of
Josephson vortices).  The axes of the Josephson strings are confined to the
insulating regions between superconducting layers.  As a consequence of the
Josephson coupling, the radial component of the magnetic field is screened on
the length scale of $\lambda_c$ by the induced Josephson currents, which flow
perpendicular to the layers.  Although there is little perturbation of the
original field distribution when $x_0 < \lambda_J$, the Josephson length, this 
situation is
altered when $x_0 > \lambda_J$, because in this case a nonlinear Josephson core
appears along the string connecting the two pancake vortices
centered at
$(x,y,z) = (0,0,0)$ and $(x,y,z) = (x_0,0,s)$.  The Josephson-energy cost of the
Josephson string coupling the two semi-infinite stacks of pancake vortices is
approximately (taking logarithmic terms to be of order
unity)\cite{Mints00,Glazman91,Kapitulnik92,Bulaevskii92a,Bulaevskii92b,Clem93}
\begin{equation}
E_{short}(x_0) \approx (\phi_0/4 \pi)^2 x_0^2/s\lambda_c^2, x_0 <
\lambda_J = (\lambda_c/\lambda_{ab})s,
\label{Eshort}
\end{equation}
when the Josephson string is short and its core is not
fully formed.  The Josephson-energy cost is of the order of
\cite{Mints00,Bulaevskii73,Clem90,Clem91a,Glazman91,Kapitulnik92,Bulaevskii92a,
Bulaevskii92b,Clem93}
\begin{equation}
E_{long}(x_0) \approx (\phi_0/4 \pi)^2 x_0/\lambda_{ab}\lambda_c, x_0 >
\lambda_J = (\lambda_c/\lambda_{ab})s,
\label{Elong}
\end{equation} 
when the Josephson string is long and its
core is more fully formed.  However, it is not until $x_0 \gg \lambda_c$ that a
fully formed Josephson vortex (with width $2\lambda_c$ and height $2\lambda_{ab}$)
can stretch out between the upper and lower parts of the split stack of pancake
vortices.  In this case the energy cost of the Josephson string coupling the two
semi-infinite stacks of pancake vortices reduces to $(\phi_0
H_{c1,ab}/4\pi)x_0$, where $\phi_0
H_{c1,ab}/4\pi$ is the energy per unit length of an isolated Josephson vortex
parallel to the layers
and\cite{Bulaevskii73,Clem91a,Bulaevskii92b,Koshelev93}
\begin{equation}
H_{c1,ab} = \frac{\phi_0}{4 \pi \lambda_{ab} \lambda_c}
\big[\ln\big(\frac{\lambda_{ab}}{s}\big) + 1.55\big]. 
\end{equation}
is the lower critical field parallel to the layers.

In anisotropic superconductors consisting of Josephson-coupled superconducting
layers, one may always regard the vortex structure as consisting of a
superposition of 2D pancake vortices, which carry magnetic flux up through the
layers, and Josephson vortices (or strings), which carry magnetic flux parallel
to the layers but no net flux perpendicular to the layers.
In transport
experiments involving vortex motion, the voltages are given by the Josephson
relations.\cite{Josephson62} The dc voltage parallel to the layers is $V_{||} =
(h/2e)\nu_{||}$, where $\nu_{||}$ is the time-averaged rate with which 2D
pancake vortices cross a line  between the contacts, and  the dc voltage
perpendicular to the layers is proportional to $V_{\bot} =
(h/2e)\nu_{\bot}$, where $\nu_{\bot}$ is the time-averaged rate with which the
axes of Josephson vortices (or strings) cross a line between the contacts.  

When the Josephson coupling is strong, vortex lines tilted with respect
to the
$z$ (or $c$) axis can be described as tilted stacks of 2D pancake vortices or as
a tilted lattice, where
 pancakes in adjacent layers are connected by Josephson strings. Such vortices,
sometimes called  kinked vortex lines,\cite{Feinberg90a,Feinberg90b} have been
studied by numerous
authors.\cite{Koshelev93,Ivlev91,Maslov91a,Maslov91b,Feinberg92} However, when the
Josephson coupling is very weak, a magnetic field applied at a small angle
relative to the layers can produce a structure consisting of two perpendicular
interpenetrating lattices\cite{Bulaevskii92a,Huse92,Koshelev99} (called a
combined lattice\cite{Bulaevskii92a} or crossing lattices\cite{Koshelev99}): a
lattice of pancake vortices aligned nearly perpendicular to the layers and a
lattice of Josephson vortices parallel to the layers.  The interaction between
the two kinds of vortices leads to striking chain-like vortex patterns in highly
anisotropic Bi-2212, which have been observed by Bitter 
decoration\cite{Bolle91,Grigorieva95} and scanning Hall-probe
microscopy.\cite{Grigorenko02,Oral96,Grigorenko01} Both techniques reveal the
positions of 2D pancake vortices within about $\lambda_{ab}$ of the surface. As
shown by Koshelev,\cite{Koshelev03} in highly anisotropic layered superconductors
the interactions between pancake vortices and Josephson vortices  lead to
deformations of both the pancake-vortex and Josephson-vortex crystals and to
pinning of Josephson vortices by pancake vortices.

At high temperatures and applied magnetic fields, the vortex lattice
melts,\cite{Safar92,Kwok92,Cubitt93,Zeldov95,Schilling96} and this process has
even been directly visualized in
Bi-2212 by scanning Hall-probe
microscopy.\cite{Oral98,Bending99} The authors of Ref.\ \onlinecite{Bending99}
used the formalism of Sec. IV A to infer the Lindemann
parameter from the rms thermal fluctuations of pancake vortices 
vs magnetic field just below the melting transition. Much
experimental and theoretical research has been devoted to vortex-lattice
melting, and the reader is referred to reviews by Blatter et
al.\cite{Blatter94} and Brandt\cite{Brandt95} for a more complete discussion of
this topic.  

The pinning of vortices by point defects is another topic where the interactions
between pancake vortices and Josephson vortices play a key role.  This
difficult  subject is further complicated by the effects of thermal fluctuations,
especially in the high-temperature superconductors at the elevated temperatures
where potential applications are most interesting.  The reader is referred to the
above reviews\cite{Blatter94,Brandt95} and the recent paper by
Kierfeld\cite{Kierfeld04} for further details about this subject. 

\section{Summary}

In this paper I have presented  solutions that permit the calculation of  the
magnetic-field and current-density distributions generated by a single 2D pancake
vortex in an infinite stack (Sec.\ II), semi-infinite stack (Sec.\ III), or a
finite-thickness stack (Sec.\ IV) of Josephson-decoupled superconducting layers. 
I have shown in Sec.\ V  how to calculate the electromagnetic forces between two
pancake vortices, and in Sec.\ VI I have discussed some of the ways that
interlayer Josephson coupling modifies the results.

The results of this paper should be useful to those using probes (such as scanning
Hall-probe microscopy, scanning SQUID microscopy, Bitter decoration, and
magneto-optics) of the vortex-generated magnetic-field distributions  above
anisotropic high-temperature superconductors.  If the sample surface is  parallel
to the cuprate planes, these probes measure chiefly the magnetic fields
generated by pancake vortices within about $\lambda_{ab}$ of the top surface.
Although Josephson vortices (or strings) produce no net magnetic flux through the
top surface, they can produce dipole-like stray fields if they are within
$\lambda_{ab}$ of the surface.  On the other hand,  if the sample surface is
normal to the cuprate planes, such probes measure chiefly the magnetic fields
generated by Josephson vortices within about
$\lambda_c$ of the sample surface, although pancake vortices within
$\lambda_{ab}$ of the surface can produce dipole-like stray fields
outside the sample.\cite{Mints96}  

The  pancake-vortex field and current distributions given in Secs.\ II-IV also
could be useful in analyzing experiments such as Lorentz
microscopy\cite{Fanesi99,Yoshida99,Matsuda01,Kamikura02,Beleggia02,Tonomura03a,
Tonomura03b}
that probe the magnetic-field distribution throughout the sample thickness.

Since the London model is at the heart of the above
pancake-vortex calculations, the resulting theoretical field and current
distributions have unphysical singularities at the pancake-vortex core, which is
of size
$\sim \xi_{ab}$.  Such singularities should have no experimental consequences for
the above probes, which have insufficient resolution to reveal details 
at this length scale.  However, for probes of higher
resolution it may be necessary to take into account the fact that the
circulating current density reaches a maximum at $\rho \approx \xi_{ab}$
and vanishes linearly  as $\rho \rightarrow 0$, such that the
singularity of the magnetic field at the pancake-vortex core is removed. The core
effects could be treated approximately by using a vortex-core model that employs a
variational core-radius parameter $\xi_v \sim \xi_{ab}$, as in Refs.\
\onlinecite{Clem75vm,Clem92,Clem80,Clem75}.

\begin{acknowledgments}
I thank S. J. Bending, P. G. Clem, J. W. Guikema, J. Hoffman, and V. G. Kogan for
stimulating discussions and correspondence.  This manuscript has been authored in
part by Iowa State University of Science  and Technology under Contract No.\
W-7405-ENG-82 with the U.S.\ Department of  Energy.
\end{acknowledgments}

\appendix*

\section{Integrals useful for $D \ll \lambda_{||}$}

Several integrals appear in Sec.\ IV B.  All may be evaluated by starting
from\cite{Gradshteyn00,Abramowitz67} 
\begin{equation}
\int_0^\infty du
\frac{J_0(zu)}{1+u}=\frac{\pi}{2}[{\bm H}_0(z)-Y_0(z)],
\label{J0start}
\end{equation} 
where ${\bm H}_n(z)$ is the Struve function and $Y_n(z)$ is the Bessel function
of the second kind (Weber's function), differentiating with respect to
$z$, making use of recurrence relations, integrating by parts, and making use of
the properties that\cite{Gradshteyn00,Abramowitz67}
\begin{eqnarray}
\int_0^\infty du J_0(zu)&=&\int_0^\infty du J_1(zu)=\frac{1}{z},
\label{Jn} \\
\int_0^\infty du \frac{J_1(zu)}{u}&=&1.
\label{J1/u}
\end{eqnarray} 
The vector potential $a_\phi(\rho,0)$ is proportional to 
\begin{eqnarray}
\int_0^\infty du
\frac{J_1(zu)}{1+u}&=&\frac{1}{z}+1-\frac{\pi}{2}[{\bm H}_1(z)-Y_1(z)] 
\label{J1/1+u}\\
&\approx& 1, z \ll 1,
\label{form1}\\
&\approx&  \frac{1}{z}, z \gg 1,
\label{form2}
\end{eqnarray} 
where the limiting forms for $z \ll 1$ and $z \gg 1$ are obtained from
expansions given in Refs.\ \onlinecite{Gradshteyn00} and
\onlinecite{Abramowitz67}.  However, Eq.\ (\ref{form1}) may be obtained more
simply by noting that,  because of the properties of
$J_1(uz)$, the integral  when $z \ll 1$ is dominated by values of $u \gg 1$, such
that $1+u$ may be replaced by
$u$; the resulting integral then takes the form of Eq.\ (\ref{J1/u}). 
Similarly, Eq.\ (\ref{form2}) may be obtained by noting that when  $z \gg
1$ the integral is dominated by  values of $u \ll 1$, such that $1+u$ may be
replaced by $1$; the resulting integral may be evaluated using Eq.\ (\ref{Jn}).
The limiting forms of the following integrals also may be obtained in a similar
fashion.

The magnetic field component $b_z(\rho,0)$ is proportional to
\begin{eqnarray}
\int_0^\infty du
\frac{uJ_0(zu)}{1+u}&=&\frac{1}{z}\int_0^\infty du\frac{uJ_1(zu)}{(1+u)^2}\\
&=&\frac{1}{z^2}\int_0^\infty du\frac{(1-u)J_0(zu)}{(1+u)^3} \\
&=&\frac{1}{z}-\frac{\pi}{2}[{\bm H}_0(z)-Y_0(z)] 
\label{uJ0/1+u}\\
&\approx& \frac{1}{z}, z \ll 1,
\label{form3}\\
&\approx&  \frac{1}{z^3}, z \gg 1,
\label{form4}
\end{eqnarray} 
and the net sheet current $K_D(\rho)$ is proportional to 
\begin{eqnarray}
\int_0^\infty du
\frac{uJ_1(zu)}{1+u}&=&\frac{1}{z}\int_0^\infty du\frac{J_0(zu)}{(1+u)^2} \\
&=&\frac{\pi}{2}[{\bm H}_1(z)-Y_1(z)] 
\label{uJ1/1+u}\\
&\approx& \frac{1}{z}, z \ll 1,
\label{form5}\\
&\approx&  \frac{1}{z^2}, z \gg 1.
\label{form6}
\end{eqnarray}

\end{document}